\documentclass[conference]{IEEEtran}
\IEEEoverridecommandlockouts
\usepackage{cite}
\usepackage{amsmath,amssymb,amsfonts}
\usepackage{algorithm}
\usepackage{algpseudocode}
\usepackage{graphicx}
\usepackage{textcomp}
\usepackage{xcolor}
\usepackage{color}
\usepackage{soul}
\usepackage{booktabs}
\usepackage{url}
\usepackage{graphicx}
\usepackage{subcaption}
\usepackage{float}

\begin{document}

\title{DP-FedLoRA: Privacy-Enhanced Federated Fine-Tuning for On-Device Large Language Models}


\author{\IEEEauthorblockN{Honghui Xu\IEEEauthorrefmark{1},
Shiva Shrestha\IEEEauthorrefmark{1},
Wei Chen\IEEEauthorrefmark{2},
Zhiyuan Li\IEEEauthorrefmark{2}, 
Zhipeng Cai\IEEEauthorrefmark{3}}
\IEEEauthorblockA{\IEEEauthorrefmark{1}Department of Information Technology, Kennesaw State University, Marietta, USA \\ Email: hxu10@kennesaw.edu, sshres16@students.kennesaw.edu}
\IEEEauthorblockA{\IEEEauthorrefmark{2}Nexa AI, Cupertino, USA. Email: \{alexchen, zack\}@nexa.ai}
\IEEEauthorblockA{\IEEEauthorrefmark{3}Department of Computer Science, Georgia State University, Atlanta, USA. Email: zcai@gsu.edu}
}





\maketitle

\begin{abstract}

As on-device large language model (LLM) systems become increasingly prevalent, federated fine-tuning enables advanced language understanding and generation directly on edge devices; however, it also involves processing sensitive, user-specific data, raising significant privacy concerns within the federated learning framework. To address these challenges, we propose DP-FedLoRA, a privacy-enhanced federated fine-tuning framework that integrates LoRA-based adaptation with differential privacy in a communication-efficient setting. Each client locally clips and perturbs its LoRA matrices using Gaussian noise to satisfy ($\epsilon$, $\delta$)-differential privacy. We further provide a theoretical analysis demonstrating the unbiased nature of the updates and deriving bounds on the variance introduced by noise, offering practical guidance for privacy-budget calibration. Experimental results across mainstream benchmarks show that DP-FedLoRA delivers competitive performance while offering strong privacy guarantees, paving the way for scalable and privacy-preserving LLM deployment in on-device environments.

\end{abstract}

\begin{IEEEkeywords}
Large Language Model, Differential Privacy, Federated Fine-tuning
\end{IEEEkeywords}

\IEEEpeerreviewmaketitle

\section{Introduction}\label{sec:introduction}

The increasing deployment of on-device large language models (LLMs) has brought powerful language understanding and generation capabilities directly to edge devices~\cite{zheng2025review,xu2024device}.
To adapt these large models to diverse user environments and device-specific tasks, federated fine-tuning has emerged as a popular solution~\cite{bian2025survey,amini2025distributed}.
Federated learning (FL) enables decentralized model adaptation by allowing edge devices to train on their local data while only sharing model updates~\cite{beltran2023decentralized,bellavista2021decentralised}.
Specifically, combined with parameter-efficient fine-tuning (PEFT) techniques, this federated learning framework allows even resource-constrained devices to fine-tune powerful LLMs collaboratively~\cite{yan2025federated,bian2025survey}.

However, this federated training paradigm introduces serious privacy risks, even though raw data never leaves the device. 
In particular, membership inference attacks (MIAs) have shown that adversaries, especially a semi-honest central server, can exploit the shared model updates to infer whether specific user data was used in training~\cite{nguyen2023active,zhang2021survey}. 
These vulnerabilities pose significant threats to user confidentiality, especially when dealing with private text, medical records, or behavioral logs processed by on-device LLMs~\cite{kibriya2024privacy,ali2025understanding}.

To bridge this gap, we propose a privacy-enhanced federated fine-tuning framework designed specifically for on-device LLM ecosystems. 
Our approach integrates Low-Rank Adaptation (LoRA), a parameter-efficient fine-tuning technique, with differential privacy (DP) in a communication-efficient federated learning setup. 
We introduce a novel algorithm, DP-FedLoRA, which enables each edge client to locally fine-tune a low-rank adaptation of the global LLM using private data, while preserving privacy through calibrated noise injection and norm clipping. 
These privacy-enhanced updates are then securely aggregated at a central server using a structured stacking mechanism that reconstructs a global adaptation layer without exposing individual client contributions, which ensures privacy protection against membership inference attacks on-device LLMs.

Beyond practical design, we provide a theoretical analysis of the privacy-utility trade-off introduced by Gaussian noise in our DP-FedLoRA framework. 
To be specific, we show that the injected noise introduces no bias in expectation but contributes additional variance, which can be bounded analytically in terms of the model size and noise scale. 
These bounds offer explicit guidelines for tuning the privacy parameters without severely degrading model performance.
Finally, we validate our proposed method through comprehensive experiments on real-world LLM benchmarks. 
Our results demonstrate that DP-FedLoRA delivers performance comparable to existing federated fine-tuning methods, while providing strong privacy guarantees across LLM-enabled edge devices.
To sum up, the key contributions of this work are summarized as follows:
\begin{itemize}
    \item We propose the first privacy-enhanced federated fine-tuning framework for on-device LLM ecosystems.

    \item We design a noise-injected aggregation mechanism that preserves the LoRA structure and accommodates heterogeneous clients with varying adaptation ranks.

    \item We provide a theoretical analysis of the expectation and variance of model updates under the Gaussian mechanism in DP, offering practical guidance for balancing privacy guarantees and learning performance.

    \item Comprehensive experiments are conducted across mainstream benchmarks to validate the effectiveness of our DP-FedLoRA.
\end{itemize}

The remainder of this paper is organized as follows. We review related work in Section~\ref{sec:related_work} and present the necessary preliminaries in Section~\ref{sec:preliminary}. The threat model is described in Section~\ref{sec:threat_modal}, and the implementation details of our proposed DP-FedLoRA framework are provided in Section~\ref{sec:methodology}. We then present the theoretical analysis in Section~\ref{sec:analysis_DP_FedLoRA}. Experimental results are discussed in Section~\ref{sec:experiment}, and we conclude the paper in Section~\ref{sec:conclusion}.

\section{Related Work}\label{sec:related_work}

In this section, we review related work on mainstream parameter-efficient fine-tuning techniques for LLMs, as well as recent advances in federated fine-tuning of LLMs.

\subsection{Parameter-Efficient Fine-tuning of LLMs}\label{subsec:peft_LLMs}

LLMs pose significant challenges in fine-tuning due to their immense size and resource requirements. 
To address this, parameter-efficient fine-tuning (PEFT) methods~\cite{xin2024parameter,xu2023parameter} have emerged as practical alternatives, enabling efficient adaptation with minimal trainable parameters.
(i) Early approaches like BitFit~\cite{zaken2021bitfit} focus on tuning only bias terms, offering comparable performance to full fine-tuning with drastically reduced computation. 
(ii) Adapter tuning inserts small trainable modules between transformer layers, allowing efficient task adaptation without modifying the base model~\cite{he2021effectiveness,siddiqui2025comparative}. 
LoRA (Low-Rank Adaptation)~\cite{hu2022lora,wang2024lora} further improves efficiency by decomposing weight updates into the product of two low-rank matrices, enabling fine-tuning with reduced memory overhead. Its flexibility and strong performance have led to widespread adoption and motivated variants such as AdaLoRA, which dynamically adjusts the parameter budget based on training needs.
(iii) Other recent efforts explore optimizing LoRA across dimentions such as layer selection~\cite{gao2024higher}, initialization~\cite{hayou2024impact}, and merging strategies~\cite{chen2024iteris}.
Our work adopts LoRA as the default PEFT method due to its simplicity, strong empirical performance across diverse downstream tasks, and compatibility with distributed training frameworks.

\subsection{Federated Fine-Tuning of LLMs}\label{subsec:fed_fine_tune_LLMs}

Federated fine-tuning of LLMs offers a promising solution for preserving data privacy while enabling collaborative learning across decentralized datasets, particularly in on-device LLM systems where heterogeneous edge devices~\cite{bian2025survey,amini2025distributed}.
(i) Early works such as FedIT integrated LoRA-based PEFT into the federated learning framework, demonstrating its practicality~\cite{wang2024flora}. However, its limited support for heterogeneous LoRA configurations restricts scalability in real-world deployments. 
(ii) Subsequent approaches attempted to address this by zero-padding LoRA modules~\cite{xu2023demodulation,Chen_Tavallaie_Nazemi_Zomaya_2024}, incurring extra computation and communication overhead, and by separately averaging LoRA’s decomposed matrices, which introduced aggregation noise. 
(iii) Other research has improved LoRA’s utility in FL through sparse initialization~\cite{10666083}, SVD-based heterogeneity mitigation~\cite{chen2024autorankmcdabasedrank}, and communication-efficient optimization~\cite{10.1145/3589334.3645702}.
While these methods improve efficiency and generalization, federated training remains vulnerable to privacy threats such as membership inference attacks~\cite{nguyen2023active,zhang2021survey}, despite keeping raw data on-device. Integrating differential privacy (DP) into the federated fine-tuning process is therefore a critical direction to provide protection guarantee against such leakage risks.

\section{Preliminary}\label{sec:preliminary}

Our proposed DP-FedLoRA framework builds upon the federated LoRA model. In this section, we first present the formulation of federated LoRA as a foundation before introducing our DP-FedLoRA approach.

\subsection{Low-Rank Adaptation (LoRA)}\label{subsec:lora}

LoRA fine-tunes large pre-trained models by introducing a low-tank decomposition into specific weight matrices without updating the original weights~\cite{zhao2024loraland310finetuned}.
Let $W\in \mathbf{R}^{m \times n}$ be a pre-trained weight matrix in a neural network layer.
Instead of directly updating $W$, LoRA keeps it frozen and adds a trainable low-rank matrix $\Delta W$, such that the update weight becomes $W' = W + \Delta W$.
The low-rank matrix $\Delta W$ is parameterized as $\Delta W = BA$, where $B\in \mathbf{R}^{m \times r}$ and $A \in \mathbf{R}^{r \times n}$.
This factorization drastically reduces the number of trainable parameters from $m \times r$ to $(rd + rk)$.
Finally, the fine-tuned weight $W'$ will be 
\begin{equation}
\label{eq:LoRA}
W' = W + BA.
\end{equation}
In the forward pass, for an input vector $x \in \mathbf{R}^{n}$, the transformed output becomes $y = Wx + BAx$, where $Wx$ is computed using the frozen base model weights, and $BAx$ is the low-rank adaptation term learned during fine-tuning.
This approach maintains the integrity of the pre-trained model while allowing task-specific adaptation with a small number of additional parameters. 

\subsection{Federated LoRA}\label{subsec:fed_lora}

Federated LoRA scheme provides a privacy-aware and communication-efficient mechanism for fine-tuning LLMs~\cite{wang2024flora}. 
Supposing that in a federated learning framework consisting of $K$ clients, each client holds a private dataset $\mathbf{D}_{k}$ and collaboratively fine-tunes a shared pre-trained LLM. 
The base weight matrix $W \in \mathbf{R}^{m \times n}$ in the LLM remains frozen across all clients.
Instead of updating $W$ directly, each client $k \in \{1,2,\cdots,K\}$ learns a low-rank adaptation $\Delta W_{k} = B_{k}A_{k}$, where $B_{k} \in \mathbf{R}^{m \times r_{k}}$ and $A_{k} \in \mathbf{R}^{r_{k} \times n}$.
This approach ensures minimal parameter overhead on resource-constrained edge devices.
During local training, each client computes the adapted weight as
\begin{equation}
W'_{k} = W + B_{k}A_{k},
\end{equation} where the matrices $B_{k}$ and $A_{k}$ are the parameters trained based on its local data $\mathbf{D}_{k}$.
After a predefined number of local epochs, each client sends its learned low-rank matrices $B_{k}$ and $A_{k}$ to central server.
The server performs a secure aggregation by using the stacking operation symbolized by $\bigoplus$.
Then, we can formalize the aggregation of LoRA modules. 
The sum of the products of $K$ LoRA module pairs is equivalent to the product of their stacked matrices:
\begin{equation}
\sum_{k=1}^{K} B_{k}A_{k} = (B_1\bigoplus \cdots \bigoplus B_K)(A_1\bigoplus \cdots \bigoplus A_K),
\end{equation} where $(B_1\bigoplus \cdots \bigoplus B_K)$ indicates each $B_{k}$ is horizontally stacked to the right one before it, and $(A_1\bigoplus \cdots \bigoplus A_K)$ indicates that each $A_{k}$ is vertically stacked below the preceding one.
Once aggregated, the central server broadcasts the updated $B \in \mathbf{R}^{m \times \sum_{k=1}^{K}r_{k}}$ and $A \in \mathbf{R}^{\sum_{k=1}^{K}r_{k} \times n}$ to all clients, allowing them to update their models as 
\begin{equation}
W'_{k} = W + BA,
\end{equation} where $B = (B_1\bigoplus \cdots \bigoplus B_K)$ and $A = (A_1\bigoplus \cdots \bigoplus A_K)$.
The aforementioned federated learning framework enables each client to adapt the model to its local context while contributing to a globally improved adaptation layer, making it ideal for real-time, on-device LLMs across distributed, resource-limited devices.

\section{Threat Model}\label{sec:threat_modal}

In federated learning, although raw data never leaves the client side, the exchanged model updates can still leak sensitive information. 
One critical threat is the Membership Inference Attack (MIA), where an adversary attempts to infer whether a particular data sample was used in the local training of a participating client.
We consider a semi-honest server that follows the federated protocol but may attempt to perform MIA on received updates. Specifically, in the LoRA-based federated setting, each client $k \in \{1, \dots, K\}$ computes a low-rank adaptation $\Delta W_k = B_k A_k$, where $B_k \in \mathbb{R}^{m \times r_k}$ and $A_k \in \mathbb{R}^{r_k \times n}$, based on its private dataset $D_k$. These matrices are transmitted to a central server for aggregation.
In MIA scenario, let $D_k$ and $D_k’$ be two neighboring datasets differing in at most one sample. The goal of the attacker is to distinguish whether the received update $\Delta W_k$ was generated using $D_k$ or $D_k’$. Formally, this can be described as distinguishing between the outputs of two mechanisms:
\begin{equation}
\label{eq:MIA}
    \mathcal{M}(D_k) = \Delta W_k = B_k A_k; \quad \mathcal{M}(D_k’) = \Delta W_k’ = B_k’ A_k’.
\end{equation}
An effective MIA~\cite{he2025labelonlymembershipinferenceattack,hu2025membershipinferenceattacksvisionlanguage} implies that the attacker can exploit differences in $\Delta W_k$ to infer the presence of specific data points, which violates data confidentiality and causes data privacy leakage.

\section{DP-FedLoRA}\label{sec:methodology}

In this work, to enhance privacy in federated LoRA and defend against membership inference attacks, we incorporate differential privacy into the training process, making it difficult for MIA to distinguish between the outputs of the two mechanisms defined in Eq.~\ref{eq:MIA}.
Each client $k \in \{1,2,\cdots,K\}$ holds local data $\mathbf{D}_{k}$ and independently fine-tunes the low-rank adaptation matrices $B_{k} \in \mathbf{R}^{m \times r_{k}}$ and $A_{k} \in \mathbf{R}^{r_{k} \times n}$, while keeping the base model weight matrix $W \in \mathbf{R}^{m \times n}$ frozen. 
After local training, each client obtains an update $\Delta W_{k} = B_{k}A_{k}$ which is not shared directly to preserve privacy. 
Instead, we ensure that the shared updates satisfy $(\epsilon, \delta)$-differential privacy through noise perturbation.
Before transmitting the local LoRA matrices to the server, each client clips their updates to control sensitivity and then adds Gaussian noise.
Specifically, for each client $k$, the matrices $B_{k}$ and $A_{k}$ are first individually clipped such that:
\begin{equation}
||B_{k}||_{F} \leq C_{B_{k}}, ||A_{k}||_{F} \leq C_{A_{k}},
\end{equation} where $||\cdot||_{F}$ denotes the Frobenius norm, and $C_{B_{k}},C_{A_{k}}$ are predetermined clipping thresholds.
After clipping, noise sample from isotropic Gaussian distributions is added to each matrix:
\begin{equation}
\tilde{B_{k}} = B_{k} + \mathcal{N}(0,\sigma^2_{B_{k}}\mathbf{I});
\end{equation}
\begin{equation}
\tilde{A_{k}} = A_{k} + \mathcal{N}(0,\sigma^2_{A_{k}}\mathbf{I});
\end{equation} where $\sigma^2_{B_{k}}$ and $\sigma^2_{A_{k}}$ are noise variances calibrated to the desired privacy budget $(\epsilon, \delta)$ and $\mathbf{I}$ denotes the identity matrix.

To be specific, the differential privacy guarantee is characterized by privacy budget $(\epsilon, \delta)$.
For the Gaussian mechanism, the standard result states that a mechanism with $l_{2}$-sensitivity $S$ and Gaussian noise of standard deviation $\sigma$ satisfies $(\epsilon,\delta)$-DP if 
\begin{equation}
\sigma \geq S \cdot \sqrt{2\log(1.25/\delta)} / \epsilon.
\end{equation}
In our case, the sensitivity of each matrix is bounded by the Frobenius norm due to clipping: $S_{B_{k}} = C_{B_{k}}$ and $S_{A_{k}} = C_{A_{k}}$.
Therefore, the noise levels must 
\begin{equation}
\sigma_{B_{k}} \geq C_{B_{k}} \cdot \sqrt{2\log(1.25/\delta)} / \epsilon_{B_{k}};
\end{equation}
\begin{equation}
\sigma_{A_{k}} \geq C_{A_{k}} \cdot \sqrt{2\log(1.25/\delta)} / \epsilon_{A_{k}};
\end{equation} where $\epsilon_{B_{k}}$ and $\epsilon_{A_{k}}$ are the target privacy budgets for $\tilde{B_{k}}$ and $\tilde{A_{k}}$, respectively.

These perturbed matrices $\tilde{B_{k}}$ and $\tilde{A_{k}}$ are then sent to the server for aggregation.
Instead of directly summing the client updates as in traditional federated averaging, we apply a structured stacking operation to combine the low-rank matrices across clients. 
More specifically, each client sends its clipped and noise-perturbed matrices $\tilde{B_{k}} \in \mathbf{R}^{m \times r_{k}}$ and $\tilde{A_{k}} \in \mathbf{R}^{r_{k} \times n}$ to the central server.
The server then constructs the global low-rank adaptation matrices $\tilde{B} \in \mathbf{R}^{m \times r}$ and $\tilde{A} \in \mathbf{R}^{r \times n}$, where $r = \sum_{k=1}^{K} r_{i}$, by horizontally stacking the $\tilde{B_{k}}$s and vertically stacking the $\tilde{A_{k}}$s as follows:
\begin{equation}
\tilde{B} = (\tilde{B_{1}} \bigoplus \cdots \bigoplus \tilde{B_{K}});
\end{equation}
\begin{equation}
\tilde{A} = (\tilde{A_{1}} \bigoplus \cdots \bigoplus \tilde{A_{K}}).
\end{equation}
This aggregation strategy preserves the low-rank structure of each client's update while allowing flexible integration of variable-rank adaptations across clients.
The resulting global update $\tilde{B}\tilde{A}$ represents a unified adaptation module that encodes contributions from all clients while maintaining privacy through injected noise. This update is broadcast back to each client, allowing them to locally reconstruct the global fine-tuned model as:
\begin{equation}
W'_{k} = W + \tilde{B}\tilde{A}.
\end{equation}
By structuring the aggregation in this way, our proposed DP-FedLoRA maintains model compatibility across clients while enabling scalable, privacy-preserving federated learning.
The pseudocode of the DP-FedLoRA algorithm is described in Algorithm~\ref{alg:DP_FedLoRA}.
\begin{algorithm}
\caption{DP-FedLoRA: Differentially Private Federated LoRA Fine-Tuning}
\begin{algorithmic}
\Require Pre-trained base model weight $W \in \mathbb{R}^{m \times n}$, number of clients $K$, noise scales $\sigma_{B_k}, \sigma_{A_k}$, clipping thresholds $C_{B_k}, C_{A_k}$, local data $\{\mathbf{D}_k\}_{k=1}^{K}$
\Ensure Global LoRA update $W' = W + \tilde{B} \tilde{A}$

\For{each client $k = 1$ to $K$ \textbf{in parallel}}
    \State Initialize $B_k \in \mathbb{R}^{m \times r_k}$, $A_k \in \mathbb{R}^{r_k \times n}$
    \State Train $B_k$, $A_k$ on local data $\mathbf{D}_k$ with $W$ frozen
    \State Clip: $B_k \leftarrow B_k \cdot \min(1, \frac{C_{B_k}}{\|B_k\|_F})$
    \State Clip: $A_k \leftarrow A_k \cdot \min(1, \frac{C_{A_k}}{\|A_k\|_F})$
    \State Add Gaussian noise: $\tilde{B_k} = B_k + \mathcal{N}(0, \sigma_{B_k}^2 \mathbf{I})$
    \State Add Gaussian noise: $\tilde{A_k} = A_k + \mathcal{N}(0, \sigma_{A_k}^2 \mathbf{I})$
    \State Send $\tilde{B_k}$ and $\tilde{A_k}$ to server
\EndFor

\State Server aggregates:
\State \quad $\tilde{B} \leftarrow (\tilde{B_1} \bigoplus \cdots \bigoplus \tilde{B_K})$ \Comment{Horizontal stacking}
\State \quad $\tilde{A} \leftarrow (\tilde{A_1} \bigoplus \cdots \bigoplus \tilde{A_K})$ \Comment{Vertical stacking}
\State Broadcast $\tilde{B}$ and $\tilde{A}$ to all clients

\For{each client $k = 1$ to $K$}
    \State Update local model: $W'_k = W + \tilde{B} \tilde{A}$
\EndFor

\end{algorithmic}
\label{alg:DP_FedLoRA}
\end{algorithm}

\section{Analysis of Noise in DP-FedLoRA}\label{sec:analysis_DP_FedLoRA}

In this section, we present an expectation and variance analysis of the impact of Gaussian noise on LoRA-based federated fine-tuning within the DP-FedLoRA framework, providing practical guidance for privacy-budget calibration.

\subsection{Expectation Analysis of Gaussian Noise in DP}\label{subsec:exp_DP_ana}

For simplicity, in DP-FedLoRA, we can treat $\tilde{B} = B + \beta$ and $\tilde{A} = A + \alpha$, where $B \in \mathbf{R}^{m \times r}$ and $A \in \mathbf{R}^{r \times n}$ are the clean low-rank components obtained from local fine-tuning, and $\beta \in \mathbf{R}^{m \times r}$, $\alpha \in \mathbf{R}^{r \times n}$ are Gaussian noise matrices sampled from isotropic distributions: 
\begin{equation}
\beta \sim \mathcal{N}(0,\sigma^2_{\beta}\mathbf{I}), \alpha \sim \mathcal{N}(0,\sigma^2_{\alpha}\mathbf{I}),
\end{equation} which are general notations for the additive noise in the aggregated LoRA matrices at the server side after receiving and stacking the clients' noise-injected updates.
Assume that $\beta$ and $\alpha$ are mutually independent and also independent of $B$ and $A$.
Then we investigate the expectation difference between the noise-injected low-rank matrix product $\tilde{B}\tilde{A}$ and the original product $BA$ in the following.

To understand how the injected noise influences the fine-tuning process, we analyze the expectation of the difference between the noise-perturbed and original updates. Specifically, we aim to compute $\mathbb{E}[\tilde{B}\tilde{A}] - \mathbb{E}[BA]$, which becomes $\mathbb{E}[(B + \beta)(A + \alpha)] - \mathbb{E}[BA]$ after substituting the perturbed matrices.

Expanding the product yields $(B + \beta)(A + \alpha) = BA + B\alpha + \beta A + \beta\alpha$. Taking the expectation of both sides, we have:
\begin{equation}
\mathbb{E}[(B + \beta)(A + \alpha)] = \mathbb{E}[BA] + \mathbb{E}[B\alpha] + \mathbb{E}[\beta A] + \mathbb{E}[\beta\alpha].
\end{equation}
Given that both $\alpha$ and $\beta$ are independent zero-mean Gaussian noise matrices and are independent of the model parameters B and A, the cross terms vanish. Specifically, 
\begin{equation}
\mathbb{E}[B\alpha] = B \cdot \mathbb{E}[\alpha] = 0;
\end{equation}
\begin{equation}
\mathbb{E}[\beta A] = \mathbb{E}[\beta] \cdot A = 0;
\end{equation}
\begin{equation}
\mathbb{E}[\beta\alpha] = 0,
\end{equation} due to the independence and zero-mean properties.
As a result, we conclude that $\mathbb{E}[\tilde{B}\tilde{A}] = \mathbb{E}[BA]$, and hence the expected update difference is zero:
\begin{equation}
\label{eq:exp_error}
\mathbb{E}[\tilde{B}\tilde{A}] - \mathbb{E}[BA] = 0.
\end{equation}
This analysis shows that the noise-injected update remains unbiased in expectation. 
Although individual updates may deviate due to the stochastic noise, the average behavior of the aggregated updates aligns with the original non-private adaptation. Thus, the differential privacy mechanism in DP-FedLoRA does not introduce bias into the global model updates, maintaining the performance of federated fine-tuning.

\subsection{Variance Analysis of Gaussian Noise in DP}\label{subsec:var_DP_ana}

While the expectation of the noise-injected low-rank adaptation $\tilde{B}\tilde{A}$ remains unbiased with respect to the original adaptation $BA$, the introduction of Gaussian noise inevitably increases the variance in the learned model update. To understand this impact, we analyze the total variance introduced by the noise.
We start by expanding the noise-injected adaptation as:
\begin{equation}
\mathrm{Var}[\tilde{B}\tilde{A}] = \mathrm{Var}[BA + B\alpha + \beta A + \beta\alpha].
\end{equation}
Assuming that the noise terms $\alpha$ and $\beta$ are independent, zero-mean Gaussian matrices and are also independent of $A$ and $B$, the cross-covariance terms in the variance expansion vanish. This simplifies the total variance to the sum of variances of individual components:
\begin{equation}
\mathrm{Var}[\tilde{B}\tilde{A}] = \mathrm{Var}[B\alpha] + \mathrm{Var}[\beta A] + \mathrm{Var}[\beta\alpha].
\end{equation}

First, we consider $\mathrm{Var}[B\alpha]$. Since $\alpha \sim \mathcal{N}(0, \sigma_\alpha^2 \mathbf{I})$ and $B$ is fixed, the resulting variance is linearly proportional to the Frobenius norm of $B$:
\begin{equation}
\mathrm{Var}[B\alpha] \leq m \sigma_\alpha^2 \cdot \|B\|_F^2.
\end{equation}
Next, for $\mathrm{Var}[\beta A]$, since $\beta \sim \mathcal{N}(0, \sigma_\beta^2 \mathbf{I})$ and $A$ is fixed, the resulting variance is similarly:
\begin{equation}
\mathrm{Var}[\beta A] \leq n \sigma_\beta^2 \cdot \|A\|_F^2.
\end{equation}
Lastly, for the composite term $\mathrm{Var}[\beta\alpha]$, the product of two independent Gaussian matrices yields a variance that depends on both noise scales and the shared inner dimension $r$ of the low-rank decomposition:
\begin{equation}
\mathrm{Var}[\beta\alpha] \leq \sigma_\beta^2 \sigma_\alpha^2 \cdot mnr.
\end{equation}
By combining these three terms, we obtain a total bound for the variance of the noise-injected update:
\begin{equation}
\label{eq:var_bound}
\mathrm{Var}[\tilde{B}\tilde{A}] \leq m \sigma_\alpha^2 \cdot \|B\|_F^2 + n \sigma_\beta^2 \cdot \|A\|_F^2 + \sigma_\beta^2 \sigma_\alpha^2 \cdot mnr.
\end{equation}

This bound highlights that the noise level $\sigma$ must be carefully chosen in relation to the Frobenius norms of the adaptation matrices $B$ and $A$, as well as the overall model dimensions, to avoid excessive update variance. Additionally, the compound error term $\sigma_\beta^2 \sigma_\alpha^2 \cdot mnr$ reveals how the structure of LoRA amplifies variance in high-rank and high-dimension models.

Based on the above analysis of updates in DP-FedLoRA, we can understand how the model’s loss is affected by using noisy updates instead of clean ones. Since the added noise is zero-mean and independent, it does not alter the expected output, keeping the average prediction unbiased. However, it introduces additional fluctuations, increasing the variance in model performance. This variance is influenced by the noise scale, model size, and the structure of the low-rank adaptation. Larger models or higher noise levels result in greater prediction variability.

\section{Experiment}\label{sec:experiment}

In this section, we first describe our experimental setup and then present comprehensive results to demonstrate the effectiveness of our proposed DP-FedLoRA framework, along with key findings derived from its evaluation. The code for all our experiments is available at: \url{https://github.com/ahahnut/DP-FedLoRA}.

\subsection{Experiment Settings}\label{subsec:exp_set}

We describe our experimental settings from three aspects: datasets, baseline methods, and training details.

\subsubsection{Datasets}\label{subsubsec:datasets}
We use the Alpaca-GPT-4 dataset for training, which is generated using GPT-4 via the Self-Instruct framework.
During training, we simulate a federated learning environment with 20 clients and randomly sample 2 clients per round. 
These selected clients collectively hold a total of 20,000 data samples. For evaluation, we consider close-ended benchmarks only. The close-ended benchmarks include MMLU (knowledge)~\cite{wang2024mmluprorobustchallengingmultitask}, BBH (reasoning)~\cite{kazemi2025bigbenchextrahard}, and CRASS (counterfactual reasoning)~\cite{frohberg2022crassnoveldataset}.

\subsubsection{Baselines}\label{subsubsec:baselines}

To gain deeper insights into the performance of existing federated learning (FL) baselines in the context of LLMs and to establish a more comprehensive evaluation framework, we implement seven representative FL algorithms. 
Specifically, we integrate FedAvg~\cite{sun2021decentralizedfederatedaveraging}, FedProx~\cite{li2020federatedoptimizationheterogeneousnetworks}, SCAFFOLD~\cite{karimireddy2021scaffoldstochasticcontrolledaveraging}, FedAvgM~\cite{sun2023roleservermomentumfederated}, FedAdagrad~\cite{DBLP:journals/fgcs/CaoWWMWHZ25}, FedYogi~\cite{baumgart2024federatedlearningalgorithmscreated}, and FedAdam~\cite{10584508}. 
Among these, FedProx and SCAFFOLD are designed to address data heterogeneity by incorporating local model correction mechanisms. 
In contrast, FedAvgM, FedAdagrad, FedYogi, and FedAdam introduce server-side momentum or adaptive optimization to stabilize global model updates.

\subsubsection{Default Training Details}\label{subsubsec:training_details}

For setting a consistent baseline across different experimental settings, a quantized LLaMA-2-7B model is used to boost memory and computation efficiency. The federated fine-tuning process is carried out over 200 rounds of communication using a single NVIDIA A100 GPU. Every participant client of this effort undergoes 10 rounds of local updates per communication round, using the AdamW optimizer. A cosine learning rate schedule is followed through rounds, along with a linear decay of the learning rate from $5 \times 10^{-5}$ to $1 \times 10^{-6}$. The maximum length of the sequence of inputs is limited to 512 tokens, while each client has a local batch size of 16. To enable parameter-efficient fine-tuning, Low-Rank Adaptation (LoRA) is used with a rank of 32 and scaling factor $\alpha=64$.

%
\begin{figure*}[htbp]
    \centering
    \begin{subfigure}{0.48\textwidth}
        \includegraphics[width=\linewidth]{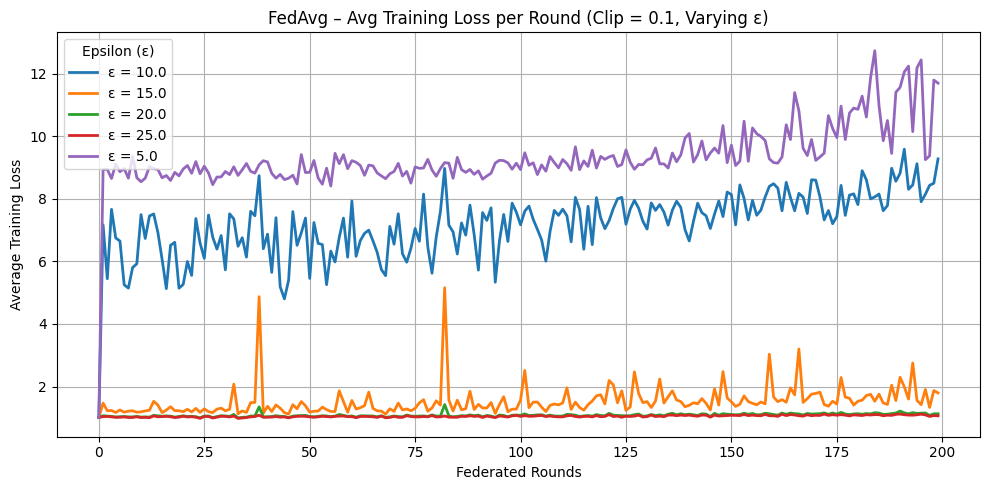}
        \caption{Varying $\varepsilon$, Clip = 0.1}
        \label{fig:ablation-epsilon-0.1}
    \end{subfigure}%
    \begin{subfigure}{0.48\textwidth}
        \includegraphics[width=\linewidth]{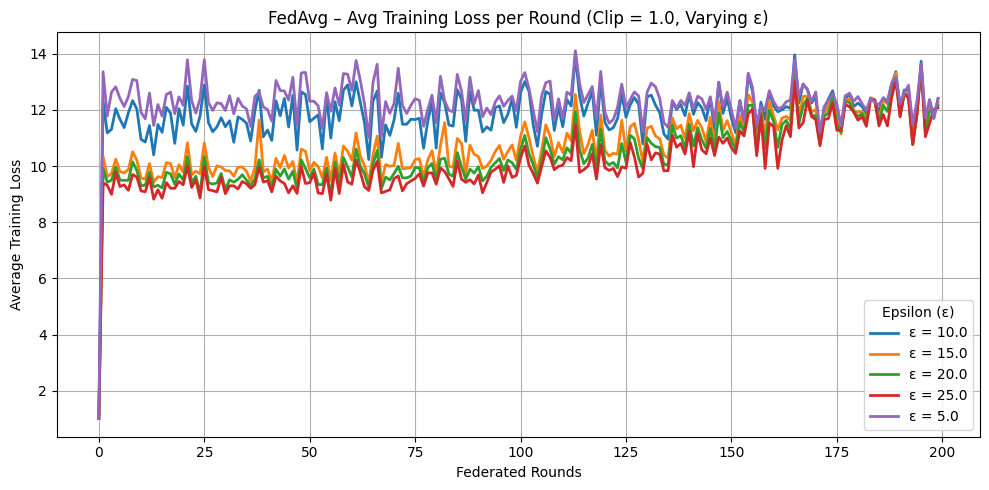}
        \caption{Varying $\varepsilon$, Clip = 1.0}
        \label{fig:ablation-epsilon-1.0}
    \end{subfigure}

    \begin{subfigure}{0.48\textwidth}
        \includegraphics[width=\linewidth]{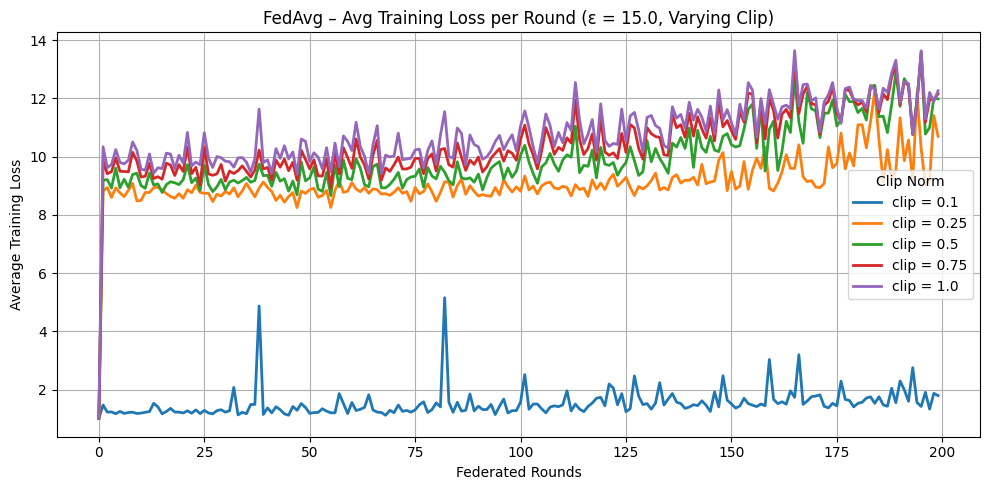}
        \caption{Varying Clip Norm, $\epsilon=15.0$}
        \label{fig:ablation-clip}
    \end{subfigure}
    \begin{subfigure}{0.48\textwidth}
        \includegraphics[width=\linewidth]{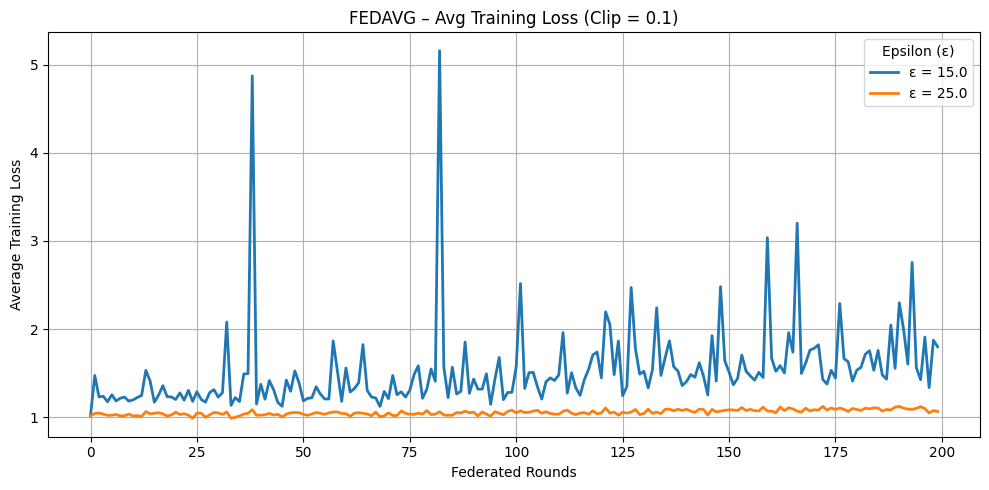}
        \caption{$\epsilon=15.0$ vs $\epsilon=25.0$ (Clip = 0.1)}
        \label{fig:ablation-compare-epsilon}
    \end{subfigure}
    \caption{Results of Ablation Study}
    \label{fig:ablation-study}
\end{figure*}

\subsection{Ablation Study}\label{subsec:abla_study}

We performed the ablation study only on the FedAvg algorithm, sweeping the differential privacy parameter $\varepsilon$ and the clip norm which is illustrated in Fig.\ref{fig:ablation-study}. 
This choice was motivated by the fact that similar phenomena emerged for other optimization algorithms and hence the choice of FedAvg was representative enough to analyze. The main goal of this study was to analyze how differential privacy hyperparameters affect model convergence and overall model stability during training. 

As shown in Fig.\ref{fig:ablation-epsilon-0.1}, when the value of $\varepsilon$ varied while keeping a minimum and fixed clipping norm of 0.1, an increment in values of $\varepsilon$ consistently produced improved convergence behavior, which is reflected by lower average losses throughout all iterations of training.
Among all settings tested, an $\varepsilon$ value of 25.0 resulted in the most stable performance curve with the lowest average loss curve, indicating that stricter privacy constraints (i.e., a larger value of $\varepsilon$) help maintain more relevant information from gradients even when there is greater differential privacy noise.
On the contrary, using stricter privacy parameters (e.g.,  $\varepsilon$= 5.0 or 10.0) resulted in significantly noisier and less stable performance measures, thus proving the trade-off between intrinsic privacy and model effectiveness.
Besides, Fig.~\ref{fig:ablation-epsilon-1.0} shows the same comparison for clip norm 1.0, which exhibits a larger average training loss across all values of $\varepsilon$ when compared to the configuration with a clip norm of 0.1.

Furthermore, we also sought to further test the impact of clipping's norm gradient by keeping $\varepsilon$ to a fixed value of 15.0 and varying clip levels from 0.1 to 1.0, as shown in Fig.~\ref{fig:ablation-clip}. 
Results indicated a significant reduction in performance to correlate with rising clipping norms, together with a parallel increase in loss values for all test conditions; namely, the lower norm—clip $= 0.1$ had improved regulated gradients and better overall convergence. As illustrated in Fig.~\ref{fig:ablation-compare-epsilon}, a careful examination between $\varepsilon$ = 15.0 and $\varepsilon$ = 25.0 for clip = 0.1 substantiated these findings: not only did the higher setting of $\varepsilon$ produce a lower final training loss, but it showed fewer and less frequent instabilities throughout training for lower $\varepsilon$. 

To sum up, the above results indicate that our setting of $\varepsilon$ = 25.0 and clip = 0.1 provides the best balance between privacy and performance under the conditions of our experimental model. Due to its improved convergence behavior and continued lowering of loss measures, we adopted this setting as the default for all subsequent differential privacy experiments under our investigation.

\begin{figure*}[htbp]
    \centering
    \begin{subfigure}[b]{0.23\textwidth}
        \includegraphics[width=\linewidth]{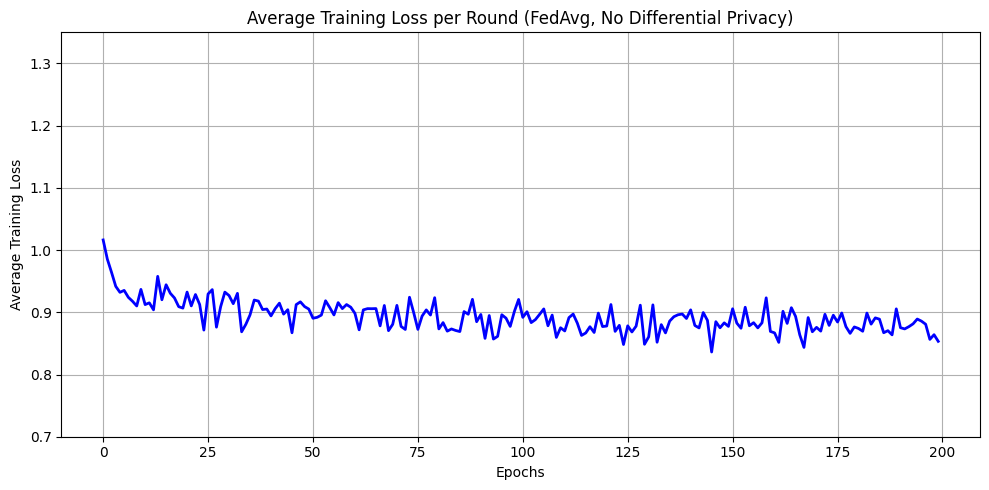}
        \caption{FedAvg}
        \label{fig:fedavg_nodp}
    \end{subfigure}
    \begin{subfigure}[b]{0.23\textwidth}
        \includegraphics[width=\linewidth]{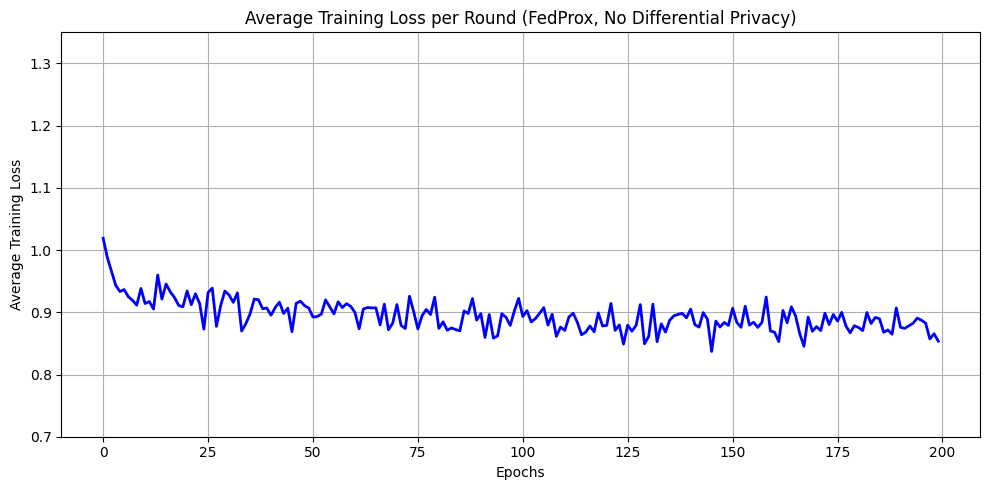}
        \caption{FedProx}
        \label{fig:fedprox_nodp}
    \end{subfigure}
    \begin{subfigure}[b]{0.23\textwidth}
        \includegraphics[width=\linewidth]{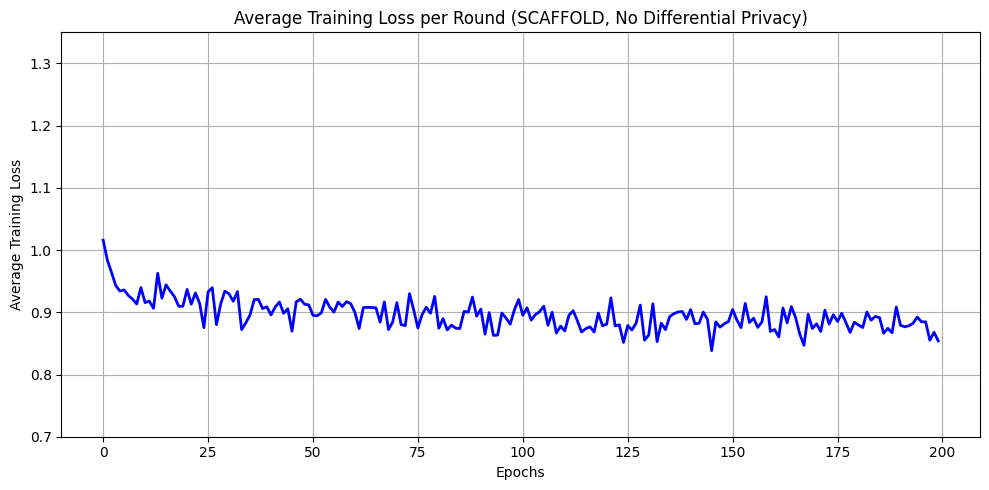}
        \caption{SCAFFOLD}
        \label{fig:scaffold_nodp}
    \end{subfigure}
    \begin{subfigure}[b]{0.23\textwidth}
        \includegraphics[width=\linewidth]{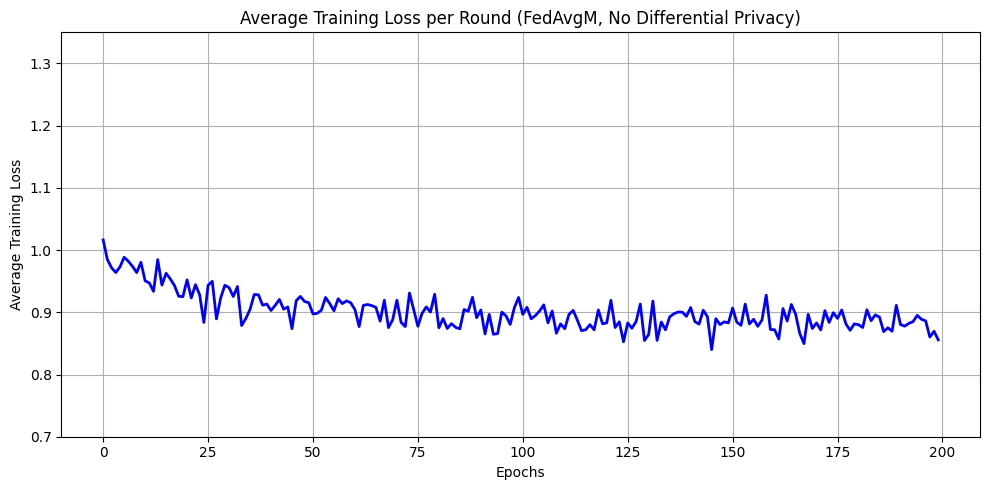}
        \caption{FedAvgM}
        \label{fig:fedavgm_nodp}
    \end{subfigure}
    
    \begin{subfigure}[b]{0.23\textwidth}
        \includegraphics[width=\linewidth]{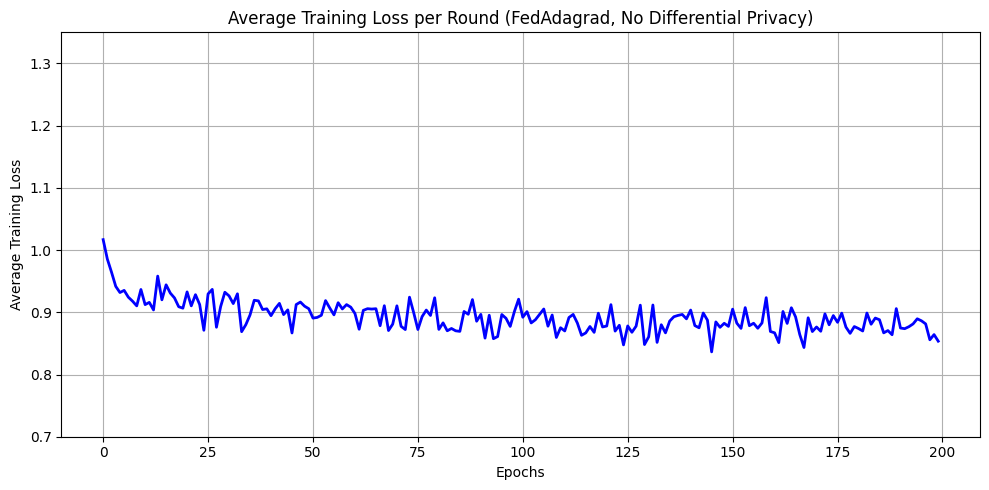}
        \caption{FedAdagrad}
        \label{fig:fedadagrad_nodp}
    \end{subfigure}
    \begin{subfigure}[b]{0.23\textwidth}
        \includegraphics[width=\linewidth]{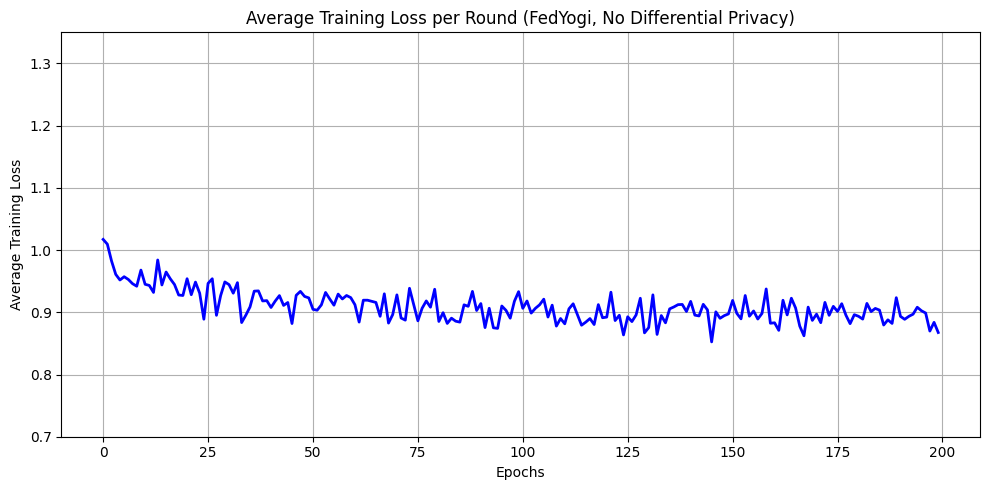}
        \caption{FedYogi}
        \label{fig:fedyogi_nodp}
    \end{subfigure}
    \begin{subfigure}[b]{0.23\textwidth}
        \includegraphics[width=\linewidth]{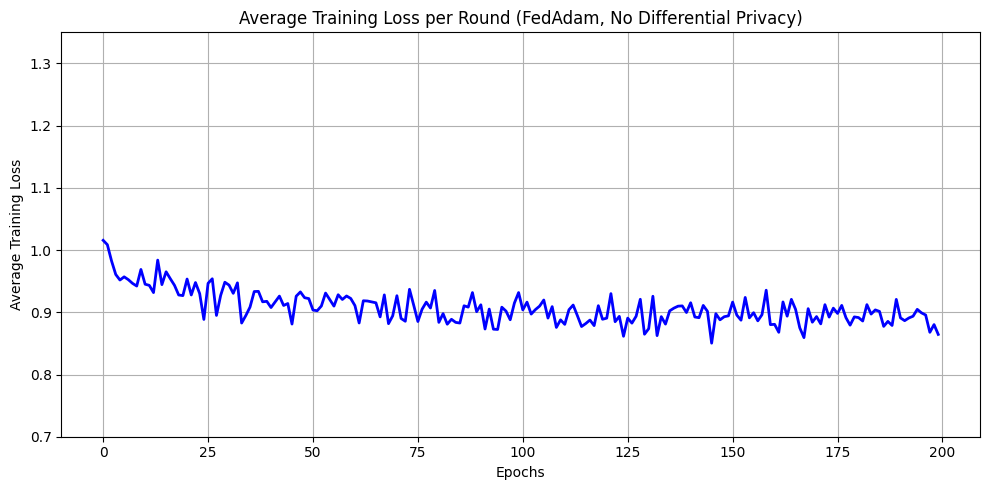}
        \caption{FedAdam}
        \label{fig:fedadam_nodp}
    \end{subfigure}
    \caption{Training Loss Curves of FedLLMs without Differential Privacy}
    \label{fig:fedllm_loss_nodp}
\end{figure*}
\begin{figure*}[htbp]
    \centering
    \begin{subfigure}[b]{0.23\textwidth}
        \includegraphics[width=\linewidth]{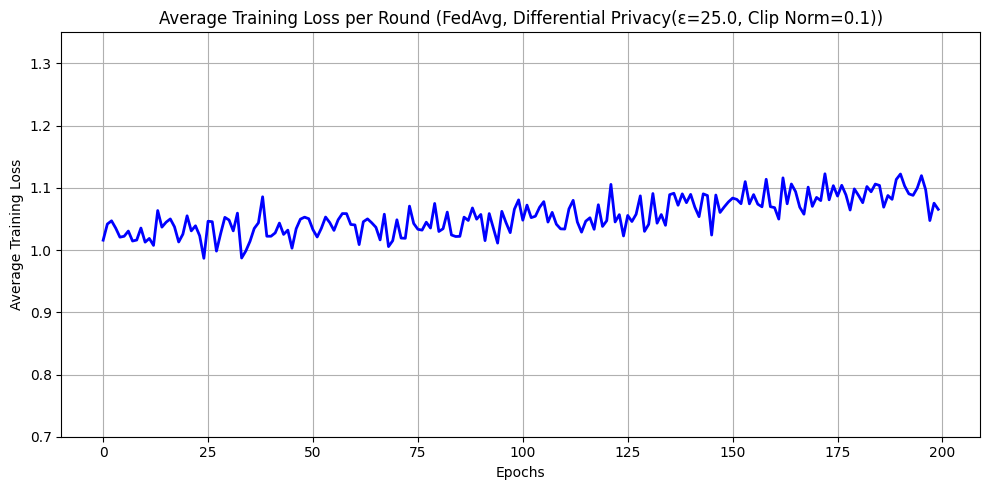}
        \caption{FedAvg}
        \label{fig:fedavg_dp}
    \end{subfigure}
    \begin{subfigure}[b]{0.23\textwidth}
        \includegraphics[width=\linewidth]{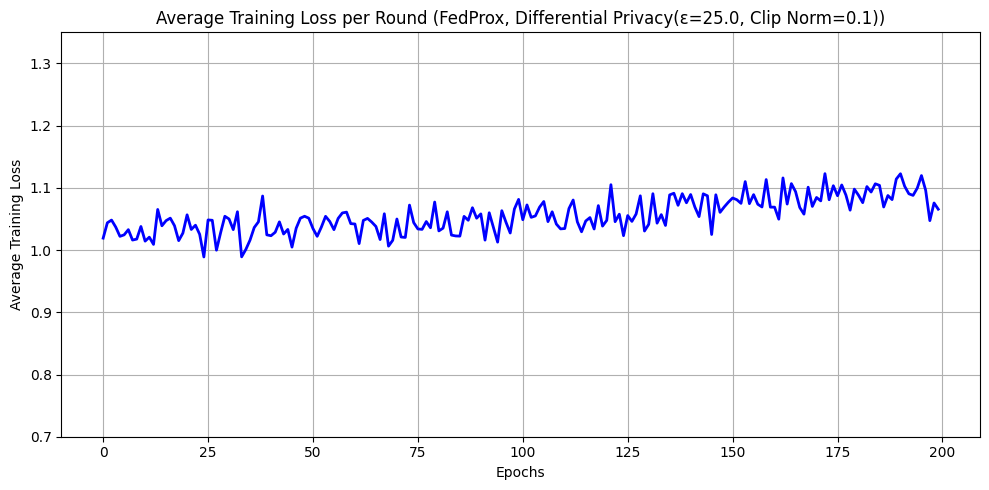}
        \caption{FedProx}
        \label{fig:fedprox_dp}
    \end{subfigure}
    \begin{subfigure}[b]{0.23\textwidth}
        \includegraphics[width=\linewidth]{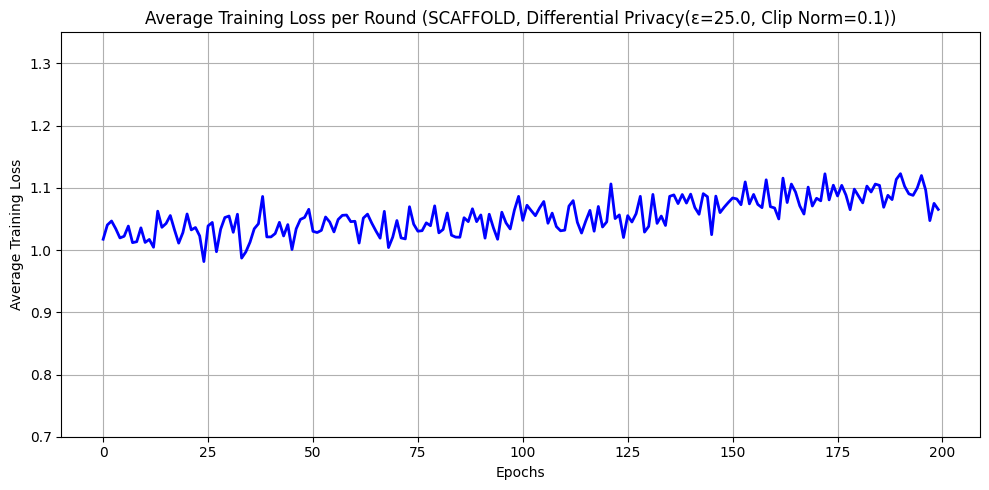}
        \caption{SCAFFOLD}
        \label{fig:scaffold_dp}
    \end{subfigure}
    \begin{subfigure}[b]{0.23\textwidth}
        \includegraphics[width=\linewidth]{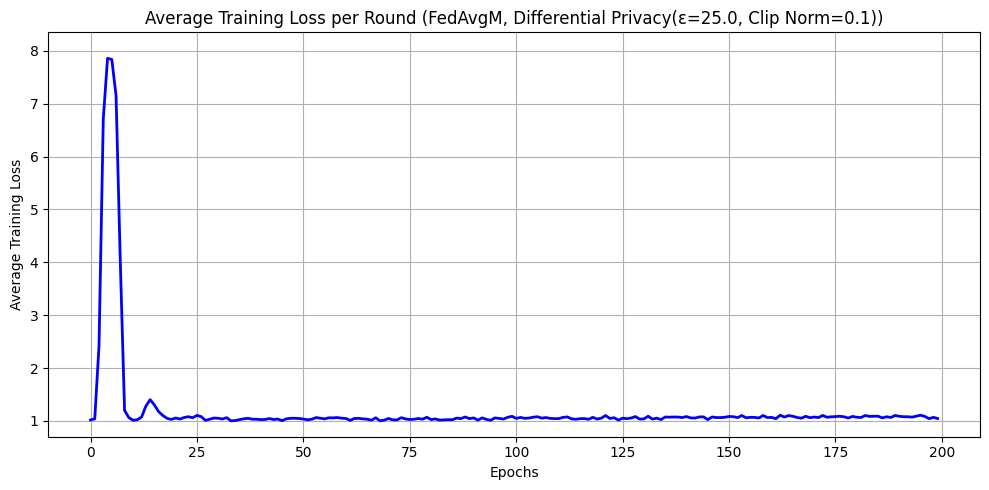}
        \caption{FedAvgM}
        \label{fig:fedavgm_dp}
    \end{subfigure}
    
    \begin{subfigure}[b]{0.23\textwidth}
        \includegraphics[width=\linewidth]{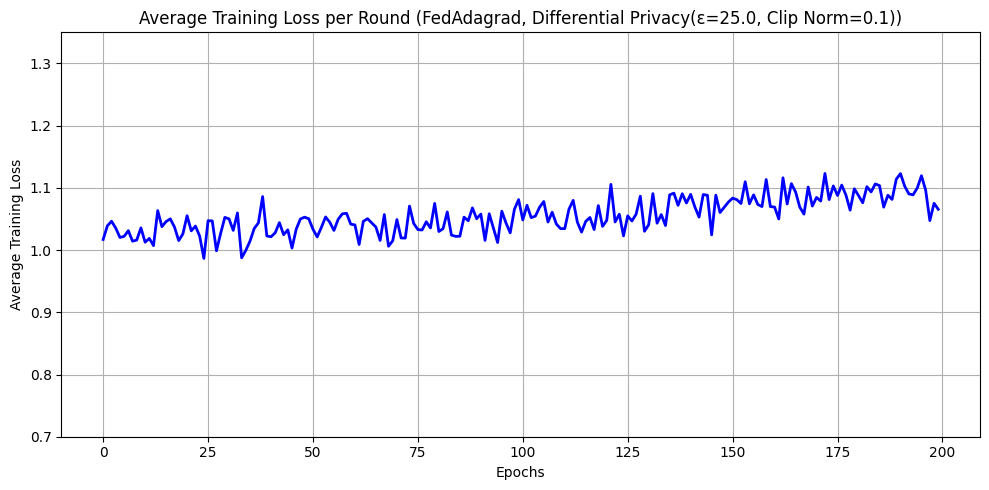}
        \caption{FedAdagrad}
        \label{fig:fedadagrad_dp}
    \end{subfigure}
    \begin{subfigure}[b]{0.23\textwidth}
        \includegraphics[width=\linewidth]{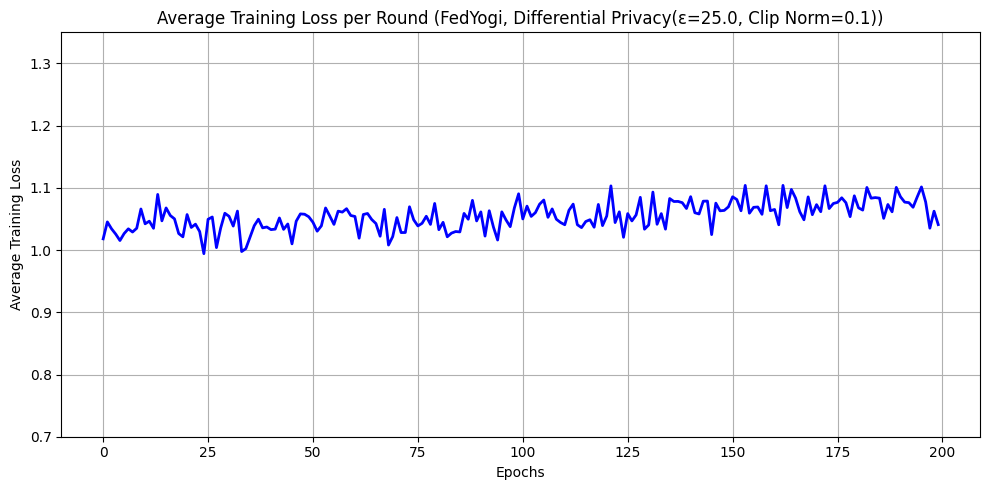}
        \caption{FedYogi}
        \label{fig:fedyogi_dp}
    \end{subfigure}
    \begin{subfigure}[b]{0.23\textwidth}
        \includegraphics[width=\linewidth]{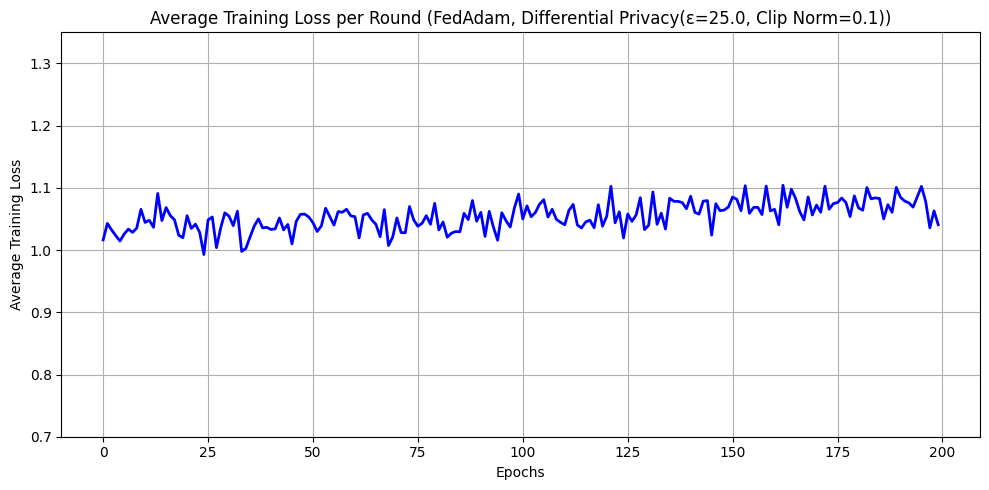}
        \caption{FedAdam}
        \label{fig:fedadam_dp}
    \end{subfigure}
    \caption{Training Loss Curves of DP-FedLLMs with Our Proposed DP-FedLoRA ($\epsilon = 25.0$, Clip = 0.1)}
    \label{fig:fedllm_loss_dp}
\end{figure*}

\subsection{Performance Evaluation: FedLLM vs. DPFedLLM}\label{subsec:DPFedLLM_vs_FedLLM}

To assess the effectiveness of our DPFedLoRA framework, we compare the training loss and downstream close-end benchmark accuracy of models trained with and without differential privacy for seven federated optimization algorithms: FedAvg, FedProx, SCAFFOLD, FedAvgM, FedAdagrad, FedYogi, and FedAdam. The differential privacy configuration is $\epsilon = 25.0$ with a clipping norm of $0.1$ for all experiments.

\subsubsection{Training Loss Curves}

Table~\ref{tab:loss_comparison} presents the average training loss after 200 communication rounds for both FedLLM and our \textbf{DPFedLLM}, which are trained using the proposed DP-FedLoRA framework, across all federated learning algorithms.
As expected, differential privacy leads to an increase in the mean training loss due to the additive noise and clipping of the gradients, which caps the learning potential. 
However, the reduction in performance is relatively moderate for all of the algorithms, especially in SCAFFOLD, indicating their robustness under privacy constraints.  
Moreover, the training loss curves in Fig.~\ref{fig:fedllm_loss_nodp} and~\ref{fig:fedllm_loss_dp} are used to further illustrate this trend, showing the progression of loss across 200 epochs without and with differential privacy, respectively. 
Due to gradient clipping and added Gaussian noise, the DP-FedLoRA models exhibit slower convergence and higher loss values overall.

As seen in Fig.~\ref{fig:fedavgm_dp}, when operating under the differential privacy setting, FedAvgM shows an elevated training loss in the first rounds—up to about 8.0—before converging sharply to about 1.0. 
Such an effect is not seen in other methods. Such an anomaly can be explained by the combination of the effect of momentum and the added differential privacy noise, coupled with aggressive clipping (clip norm = 0.1). 
Since prior gradients are accumulated by FedAvgM, the effect of noisy gradients in the first rounds can be compounded, causing divergence-like behavior. 
Even though this spike does not lead to divergence, the model eventually converges. However, this effect suggests that FedAvgM is possibly more sensitive to privacy parameters compared to other methods.

\begin{table}[htbp]
\caption{Training Loss Comparison of FedLLM and DPFedLLM after 200 Communication Rounds ($\varepsilon$ = 25.0, Clip = 0.1)}
\centering
\begin{tabular}{lcc}
\hline
\textbf{Algorithm} & \textbf{FedLLM Loss} & \textbf{DPFedLLM Loss} \\
\hline
FedAvg     & 0.8362 & 0.9868 \\
FedProx    & 0.8370 & 0.9890 \\
SCAFFOLD   & 0.8384 & 0.9815 \\
FedAvgM    & 0.8400 & 1.0003 \\
FedAdagrad & 0.8364 & 0.9866 \\
FedYogi    & 0.8524 & 0.9942 \\
FedAdam    & 0.8504 & 0.9929 \\
\hline
\end{tabular}
\label{tab:loss_comparison}
\end{table}

\subsubsection{Evaluation on Close-ended Benchmarks}

Table~\ref{tab:benchmarkComparision} summarizes the performance results, in terms of test scores, across different federated learning methods in both non-private (Non-DP) and differentially private (DP) ({\em i.e.}, our proposed DP-FedLoRA framework) setups over three prominent close-end benchmarking tasks: MMLU, BBH, and CRASS. These benchmarking tasks aim to measure a wide range of skills, from factual knowledge to reasoning abilities, and commonsense reasoning.
Differential privacy seems to have little effect on model efficacy, as indicated by the relatively small decreases in MMLU and BBH performance, with average score reductions of approximately 4–5\% across all algorithms. For example, the MMLU value for FedAvg goes from 44.64\% (Non-DP) to 42.45\% (DP), while that for BBH drops from 38.96\% to 38.52\%. this implies that private training methodologies can be used without affecting the model in scenarios that require reasoning knowledge.

In contrast, a more prominent reduction in performance is reported on the CRASS benchmark when differential privacy (DP) is introduced, suggesting that commonsense reasoning tasks are especially vulnerable to interference introduced by differential privacy. Still, this reduction is within a range (i.e., from 40.90\% to 25.00\%, i.e., FedProx), which suggests that differentially private federated learning should be improved in the counterfactual reasoning LLMs.
To conclude, our findings show our proposed DP-FedLoRA can be an effective method to fine-tune LLMs while protecting data privacy.

\begin{table}[htbp]
\centering
\caption{Evaluation Scores of Federated Learning Algorithms with and without Differential Privacy(DP)}
\label{tab:benchmarkComparision}
\begin{tabular}{lcccc}
\hline
\textbf{Algorithm} & \textbf{Privacy} & \textbf{MMLU} & \textbf{BBH} & \textbf{CRASS} \\
\hline
FedAvg      & Non-DP &  44.64 & 38.96 & 38.09 \\
FedAvg      & DP          &  42.45 & 38.52 & 22.73 \\
\hline
FedProx     & Non-DP &  44.70 & 38.79 & 40.90  \\
FedProx     & DP          &  41.92 & 37.33 & 25.00 \\
\hline
SCAFFOLD    & Non-DP &  43.68 & 39.88 & 29.54 \\
SCAFFOLD    & DP          &  41.92 & 37.70 & 25.00 \\
\hline
FedAvgM     & Non-DP &  42.07 & 38.29 & 38.64 \\
FedAvgM     & DP          &  42.57 & 37.48 & 20.45 \\
\hline
FedAdagrad  & Non-DP &  44.94 & 39.67 & 31.82 \\
FedAdagrad  & DP          &  42.58 & 37.07 & 22.28 \\
\hline
FedYogi     & Non-DP          &  44.63 & 38.52 & 34.09 \\
FedYogi     & DP &  41.57 & 37.70 & 22.73 \\
\hline
FedAdam     & Non-DP &  44.95 & 38.63  & 36.36 \\
FedAdam     & DP          &  42.35 & 38.59 & 20.45 \\
\hline
\end{tabular}
\end{table}

\begin{figure}[htbp]
    \centering
    \begin{subfigure}[b]{0.45\textwidth}
        \centering
        \includegraphics[width=\textwidth]{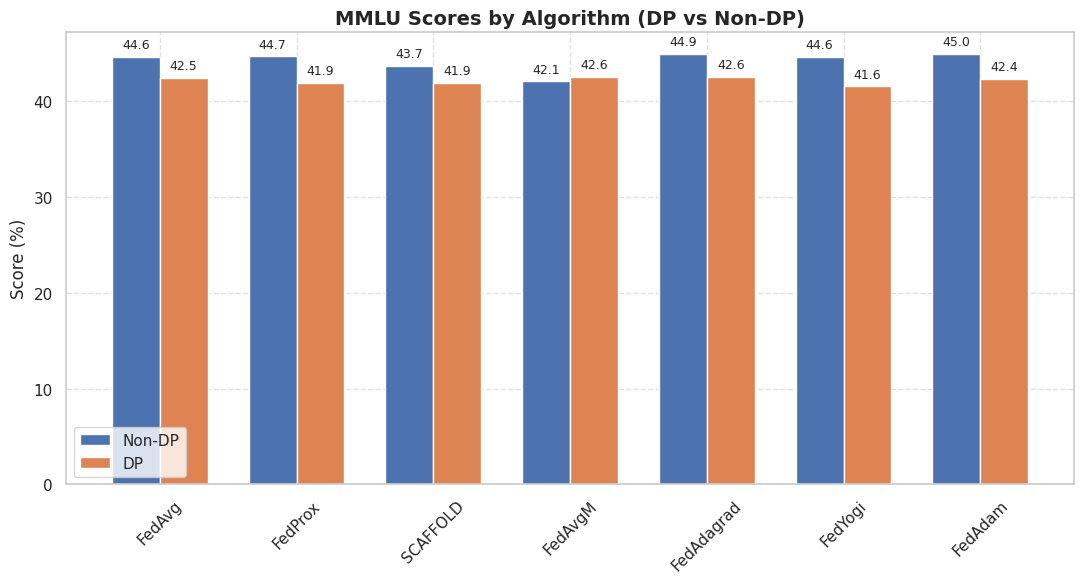}
        \caption{MMLU Score}
        \label{fig:mmlu_score}
    \end{subfigure}
    \begin{subfigure}[b]{0.45\textwidth}
        \centering
        \includegraphics[width=\textwidth]{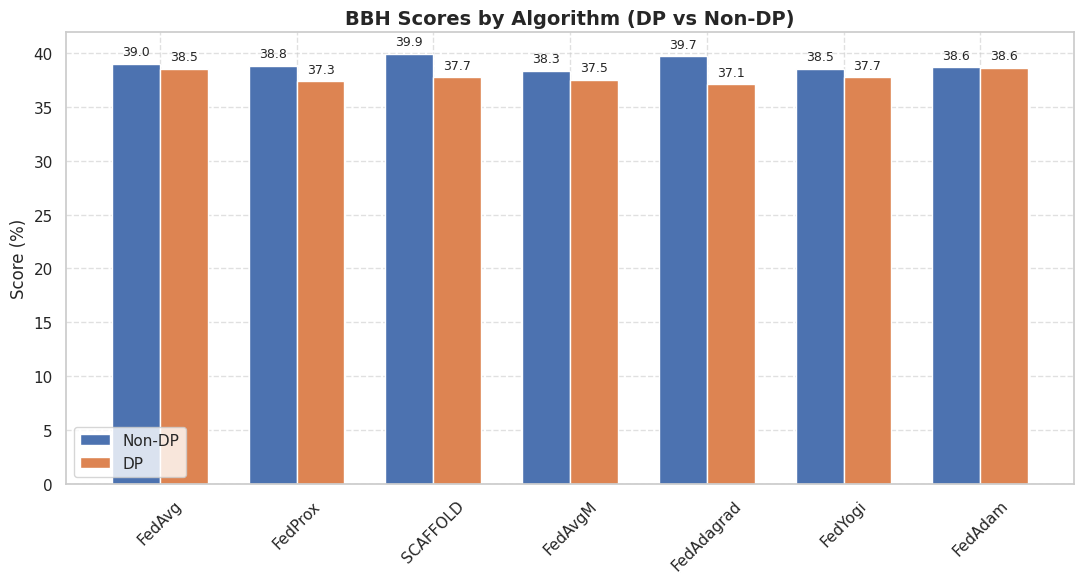}
        \caption{BBH Score}
        \label{fig:bbh_score}
    \end{subfigure}
    \begin{subfigure}[b]{0.45\textwidth}
        \centering
        \includegraphics[width=\textwidth]{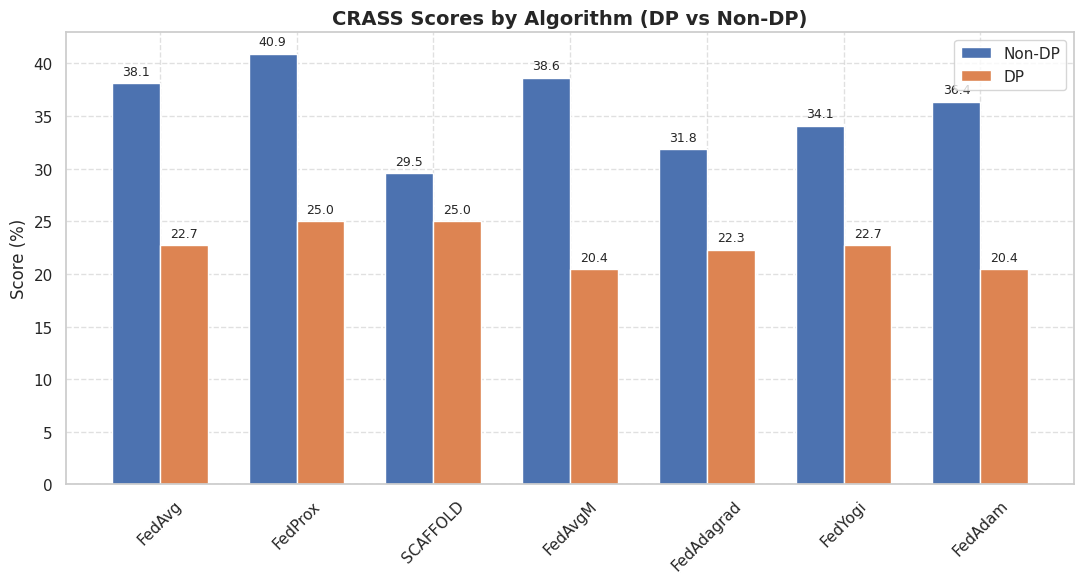}
        \caption{CRASS Score}
        \label{fig:crass_score}
    \end{subfigure}
    \caption{Performance scores of FedLLMs and DPFedLLMs for MMLU, BBH, and CRASS benchmarks}
    \label{fig:benchmark_scores}
\end{figure}

\begin{figure}[htbp]
    \centering
    \begin{subfigure}[b]{0.42\textwidth}
        \includegraphics[width=\textwidth]{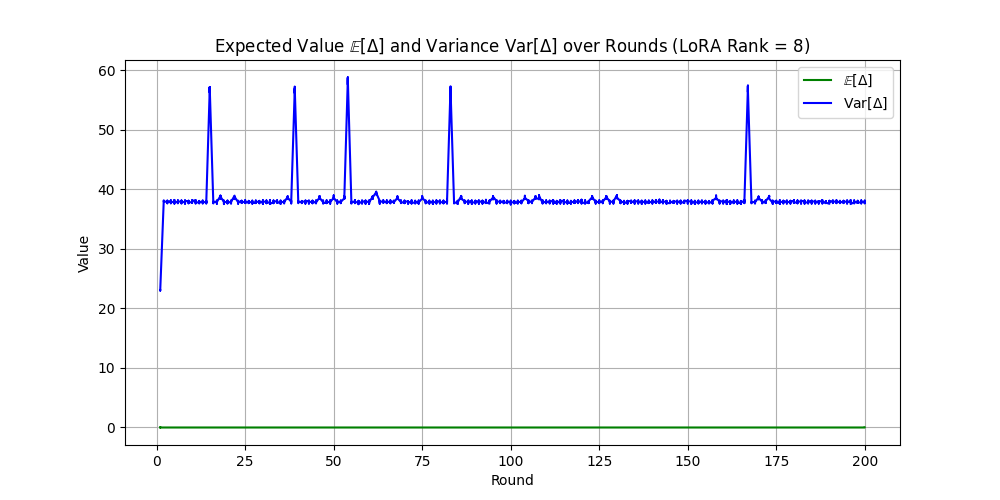}
        \caption{Rank 8}
        \label{fig:rank8}
    \end{subfigure}
    \begin{subfigure}[b]{0.42\textwidth}
        \includegraphics[width=\textwidth]{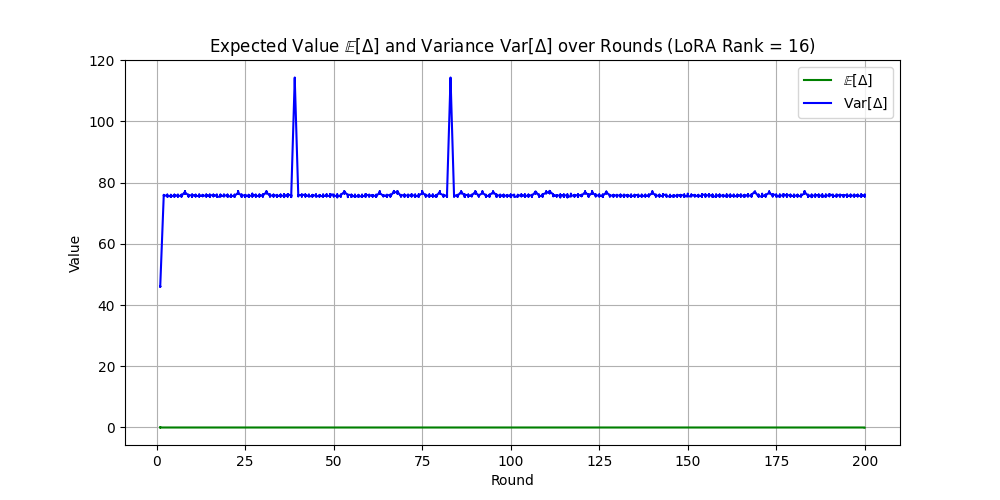}
        \caption{Rank 16}
        \label{fig:rank16}
    \end{subfigure}
    \begin{subfigure}[b]{0.42\textwidth}
        \includegraphics[width=\textwidth]{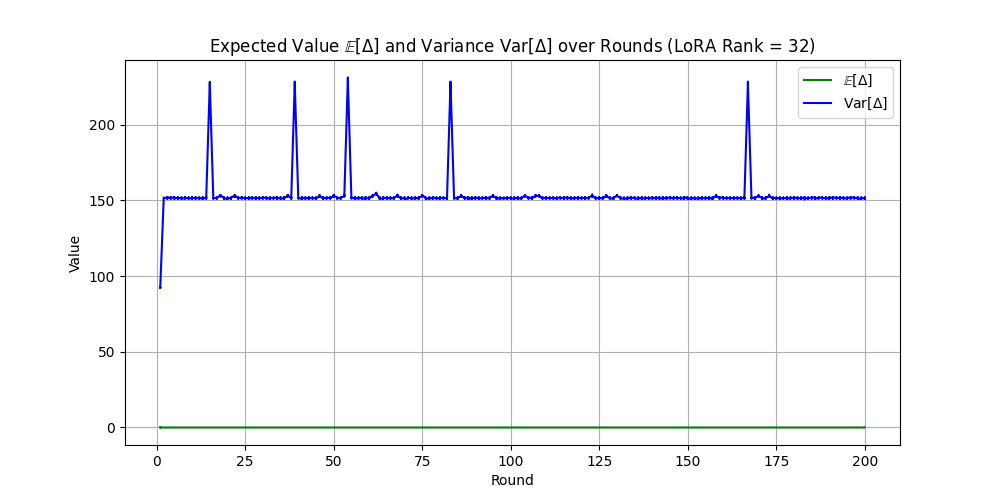}
        \caption{Rank 32}
        \label{fig:rank32}
    \end{subfigure}
    \begin{subfigure}[b]{0.42\textwidth}
        \includegraphics[width=\textwidth]{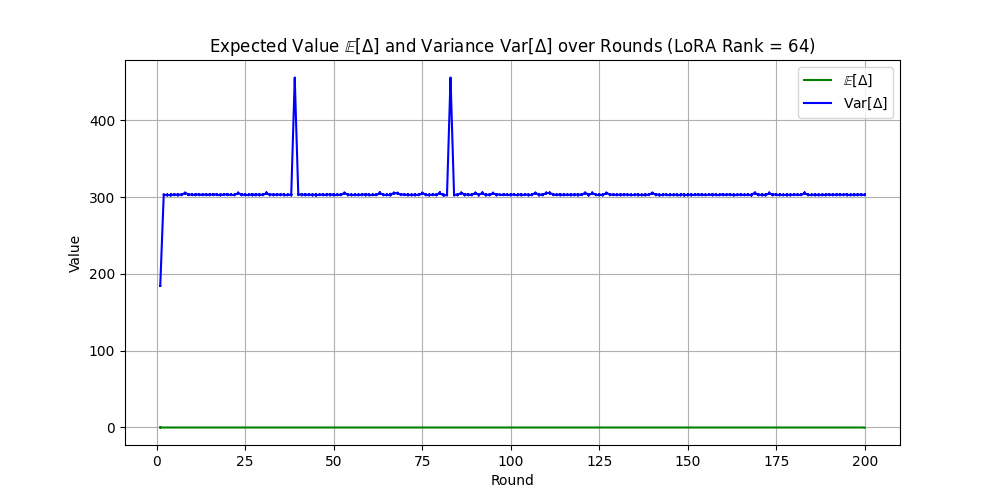}
        \caption{Rank 64}
        \label{fig:rank64}
    \end{subfigure}
    \begin{subfigure}[b]{0.42\textwidth}
        \includegraphics[width=\textwidth]{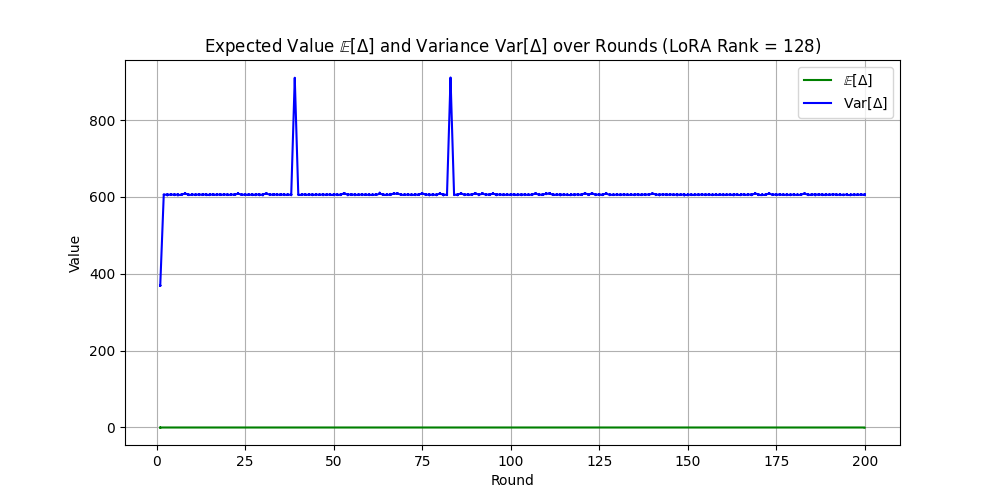}
        \caption{Rank 128}
        \label{fig:rank128}
    \end{subfigure}
    \caption{Expectation Value $\mathbb{E}[\tilde{\Delta}]$ and Variance $\mathrm{Var}[\tilde{\Delta}]$ of Noisy updates in DP-FedLoRA with Various Ranks}
    \label{fig:rank_variation}
\end{figure}

\subsection{Impact of Rank in DP-FedLoRA}\label{subsec:impact_rank}

To investigate how the Low-Rank Adaptation (LoRA) rank affects the behavior of noise introduced in our proposed DP-FedLoRA framework, experiments were performed with FedAvg as the base aggregation algorithm with a LLaMA2-7B model, having the privacy budget $\varepsilon$ constant at 25.0, the clipping norm being 0.1, and the LoRA alpha $\alpha$ being 128. 
The only tunable hyperparameter was the LoRA rank, tested at the values $\{8, 16, 32, 64, 128\}$ using the aggregation method of FedAvg. 
During these experimental processes, statistical properties of the noise-influenced updates were tracked, with a specific focus on their expectation and variance across the communication rounds.

In Fig.~\ref{fig:rank_variation}, our experimental results show that the expectation difference ($\mathbb{E}[\tilde{\Delta}]$) between the noise-injected and original updates remained near zero throughout the training process across all LoRA ranks as indicated in the Eq.\eqref{eq:exp_error}, thus establishing the evidence that the differential privacy mechanism introduced unbiased noise, which is independent of LoRA configuration. This also confirms that average update behaviors are consistent with those derived for non-private learning, thereby preserving the correctness of the federated optimization trajectory in the limit.

Besides, as illustrated in Fig.~\ref{fig:rank_variation}, the variance ($\mathrm{Var}[\tilde{\Delta}]$) of the updates showed an upward trend corresponding to the rank of LoRA. To be specific, the variances observed were around 37 at rank 8, 75 at rank 16, 150 at rank 32, 300 at rank 64, and 600 at rank 128. This regular monotonic increase in variance is consistent with the previously derived theoretical upper bound, particularly the term $\sigma_{\beta}^2 \sigma_{\alpha}^2 \cdot mnr$ from Eq.\eqref{eq:var_bound}, showing linear growth with the inner rank $r$. Therefore, higher rank adaptations naturally increase the scale of the noise, regardless of the noise scale and clipping parameter constancy.

Notably, while there is a common trend of variance growth with respect to LoRA rank, variations and occasional spikes in variance in Fig.~\ref{fig:rank_variation} were also identified throughout training across all ranks. Such spikes are identified as nonsystematic; that is, they can be caused by differences in the gradient distributions between clients, particularly when the participating clients have heterogeneous data or model states.
In federated learning setups, such non-IID scenarios where clients hold diverse and potentially skewed data distributions, can easily cause occasional spikes in both sensitivity and variability of norms, which in turn affects the amount of noise injected in each iteration even in cases where clipping mechanisms are used. 
Thus, such spikes are viewed as a natural effect of the dynamic, decentralized, and data-heterogeneous nature of federated optimization in differential privacy settings.
 
From Table~\ref{tab:convergence_rank}, we can also conclude that while our proposed DP-FedLoRA framework promises unbiased updates, an increase in the adaptation rank significantly raises the noise variance.
This trade-off between expressiveness and stability requires careful adjustment to maintain an efficient balance between privacy and utility in practical applications.

\begin{table}[ht]
\centering
\caption{Converged Expectation Value $\mathbb{E}[\tilde{\Delta}]$ and Variance $\mathrm{Var}[\tilde{\Delta}]$ of Noisy updates in DP-FedLoRA with Various Ranks}
\begin{tabular}{ccc}
\toprule
\textbf{LoRA Rank} & \textbf{Expectation ($\mathbb{E}[\tilde{\Delta}]$)} & \textbf{Variance ($\mathrm{Var}[\tilde{\Delta}]$)} \\
\midrule
8 & $-2.38 \times 10^{-8}$ & 37.88  \\
16 & $-4.27 \times 10^{-9}$ & 75.75 \\
32 & $1.04 \times 10^{-8}$ & 151.47 \\
64 & $1.26 \times 10^{-8}$  & 302.92  \\
128 & $-1.36 \times 10^{-8}$ & 605.82 \\
\bottomrule
\end{tabular}
\label{tab:convergence_rank}
\end{table}

\subsection{Impact of Parameter Size in DP-FedLoRA}

To evaluate the effect of model size on the statistical properties of noise during federated fine-tuning, we performed a comparative study of both the LLaMA-2-7B and LLaMA-2-13B under the same training setup. This setup used the FedAvg aggregation method, had a constant LoRA rank of 32, a privacy budget of $\varepsilon = 25.0$, a clipping norm of 0.1, and a LoRA scaling factor $\alpha = 128$. All other settings remained unchanged.

\begin{figure}[htbp]
    \centering
    \begin{subfigure}[b]{0.45\textwidth}
        \includegraphics[width=\textwidth]{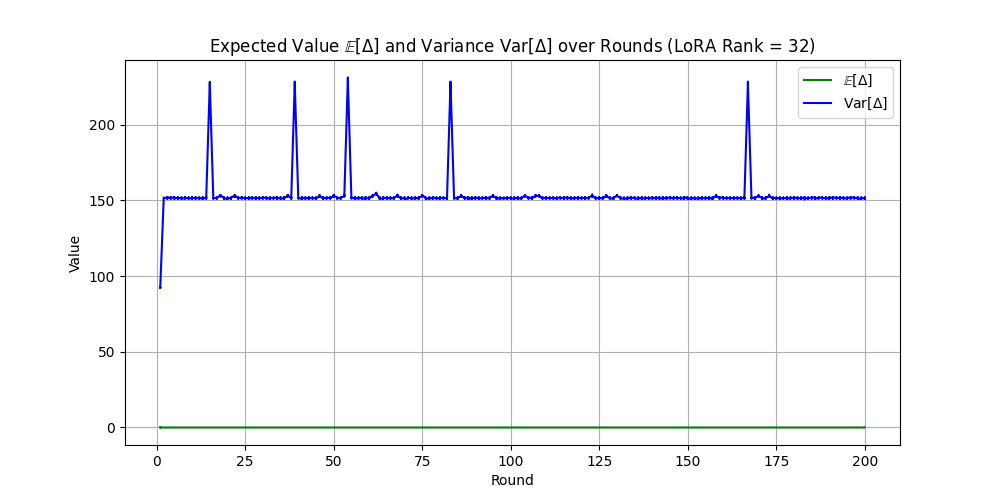}
        \caption{LLaMA-2-7B (DP-FedLoRA with Rank = 32)}
    \end{subfigure}
    \hfill
    \begin{subfigure}[b]{0.45\textwidth}
        \includegraphics[width=\textwidth]{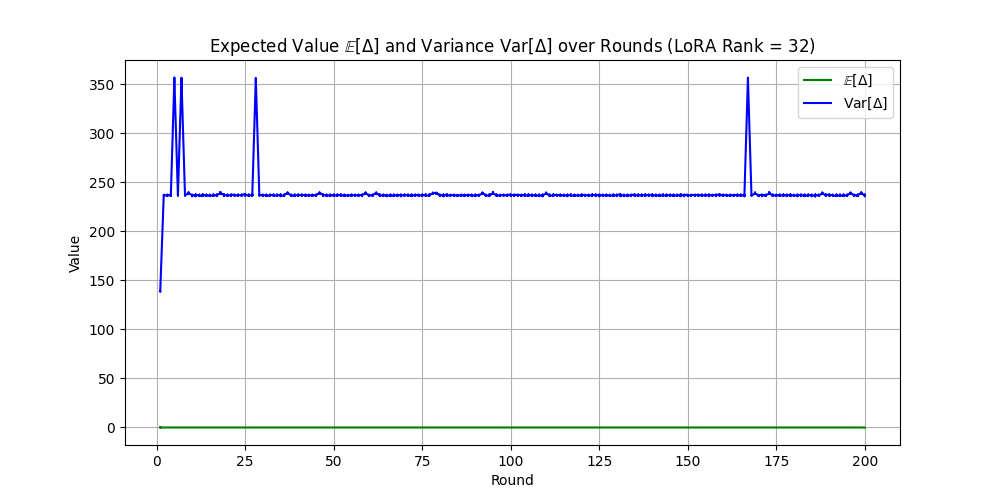}
        \caption{LLaMA-2-13B (DP-FedLoRA with Rank = 32)}
    \end{subfigure}
    \caption{Expectation $\mathbb{E}[\tilde{\Delta}]$ and Variance $\mathrm{Var}[\tilde{\Delta}]$ of Noisy Updates in Our DP-FedLoRA over training rounds with Different Base Models and Fixed LoRA Rank}
    \label{fig:model_size_variation}
\end{figure}

As shown in Fig.\ref{fig:model_size_variation}, $\mathbb{E}[\tilde{\Delta}]$ values in all communication rounds indicates the addition of noise is unbiased regardless of the model size. 
However, the variance represented as $\mathrm{Var}[\tilde{\Delta}]$ shows a considerable difference. The LLaMA-2-13B model always shows a higher variance than the 7B model. This increase is consistent with the theoretical bound concerning the matrix dimensions $m$ and $n$ in the term $\sigma_{\beta}^2 \sigma_{\alpha}^2 \cdot mnr$, which argues that larger models inherently increase the noise variance while keeping the same differential privacy mechanism.

Similarly, as shown in Table~\ref{tab:ev_var_models}, we observe empirical results that confirm our theoretical findings in Section~\ref{sec:analysis_DP_FedLoRA}: the expectation remains invariant to model size, while the variance increases with larger model sizes.

\begin{table}[h]
\centering
\caption{Converged Expectation $\mathbb{E}[\tilde{\Delta}]$ and Variance $\mathrm{Var}[\tilde{\Delta}]$ of Noisy Updates in Our DP-FedLoRA over training rounds with Different Base Models and Fixed LoRA Rank}
\begin{tabular}{ccc}
\hline
\textbf{Base Model} & \textbf{Expectation ($\mathbb{E}[\tilde{\Delta}]$)} & \textbf{Variance ($\mathrm{Var}[\tilde{\Delta}]$)} \\
\hline
LLaMA-2 7B  & $1.04 \times 10^{-8}$ & 151.47 \\
LLaMA-2 13B & $2.39 \times 10^{-9}$ & 237.50   \\
\hline
\end{tabular}
\label{tab:ev_var_models}
\end{table}

\section{Conclusion}\label{sec:conclusion}

In this paper, we proposed DP-FedLoRA, a privacy-enhanced federated fine-tuning framework for on-device LLMs deployed on edge devices. Our approach combines LoRA-based parameter-efficient adaptation with differential privacy to safeguard sensitive local data while preserving the performance of federated LLMs. To be specific, we introduced a structured noise injection and aggregation mechanism that enforces differential privacy on client updates and supports heterogeneous adaptation ranks. Additionally, we provided a theoretical analysis demonstrating the unbiased nature and bounded variance of noise-injected updates, offering practical guidance for privacy-budget calibration in federated fine-tuning. Finally, extensive experiments on real-world LLM benchmarks validate that DP-FedLoRA achieves strong privacy guarantees with minimal performance loss, presenting a scalable and effective solution for privacy-preserving LLM deployment in edge devices.

\section*{Acknowledge}\label{sec:ack}
This work was partly supported by the National Science Foundation of U.S. (2416872, 2244219, 2315596, 2146497).

%
\bibliographystyle{IEEEtran}
\bibliography{LLM_DP_Fed}

\end{document}